%% file: preprint.tex
\documentclass{article}
\RequirePackage{filecontents}
\PassOptionsToPackage{dvipsnames}{xcolor}
\RequirePackage{comgipp}
\RequirePackage{amsfonts}
\RequirePackage{amsmath}
\RequirePackage{authblk}

\newcommand{\theReferenceText}{\fullcite{SchubotzSTKBG20}}
\input{header.tex}
\input{authblk.tex}
\begin{document}
  \maketitle
  \thispagestyle{firststyle}
  \begin{abstract}
    \input{abstract.tex}
  \end{abstract}
\input{mainmatter.tex}
\end{document}

%% file: authblk.tex
\author[1,2]{Moritz Schubotz}
\author[3]{Philipp Scharpf}
\author[1]{Olaf Teschke}
\author[1]{Andreas K\"{u}hnemund}
\author[2,3]{Corinna Breitinger}
\author[2,3]{Bela Gipp}
\affil[1]{FIZ-Karlsruhe, Germany (\{first.last\}@fiz-karlsruhe.de)}
\affil[2]{University of Wuppertal, Germany ({\{last\}@uni-wuppertal.de})}
\affil[3]{University of Konstanz, Germany (\{first.last\}@uni-konstanz.de)}

%% file: abstract.tex
Authors of research papers in the fields of mathematics, and other math-heavy disciplines commonly employ the Mathematics Subject
Classification (MSC) scheme to search for relevant literature.
The MSC is a hierarchical alphanumerical classification scheme that allows librarians to specify one or multiple codes for publications.
Digital Libraries in Mathematics, as well as reviewing services, such as zbMATH and Mathematical Reviews (MR) rely on these MSC labels in their workflows to organize the abstracting and reviewing process.
Especially, the coarse-grained classification determines the subject editor who is responsible for the actual reviewing process.

In this paper, we investigate the feasibility of automatically assigning a coarse-grained primary classification using the MSC scheme, by regarding the problem as a multi class classification machine learning task.
We find that the our method achieves an \(F_{1}\)-score of over 77\%, which is remarkably close to the agreement of zbMATH and MR (\(F_{1}\)-score of 81\%).
Moreover, we find that the method's confidence score allows for reducing the effort by 86\% compared to the manual coarse-grained classification effort while maintaining a precision of 81\% for automatically classified articles.

%% file: mainmatter.tex
\section{Introduction}\label{sec:intro}
zbMATH\footnote{\href{https://zbmath.org/}{{https://zbmath.org/}}} has classified more than 135k articles in 2019 using the Mathematics Subject Classification~(MSC) scheme~\cite{Khnemund2016}.
With more than \href{https://zbmath.org/classification/}{{6,600}} MSC codes, this classification task requires significant in-depth knowledge of various sub-fields of mathematics to determine the fitting MSC codes for each article.
In summary, the classification procedure of zbMATH and MR is two-fold.
First, all articles are pre-classified into one of  \href{https://msc2020.org}{{63}}  primary subjects spanning from general topics in mathematics (00), to integral equations (45), to mathematics education (97).
In a second step, subject editors assign fine-grained MSC codes in their area of expertise, i.a. with the aim to match potential reviewers.

The automated assignments of MSC labels has been analyzed by \citeauthor{RehurekS08}~\cite{RehurekS08} in \citeyear{RehurekS08} on the DML-CZ~\cite{SojkaR07} and NUMDAM~\cite{BoucheL17} full-text corpus.
They report a micro-averaged $F_1$ score of 81\% for their public corpus.
In \citeyear{BarthelTB13} \citeauthor{BarthelTB13} performed automated subject classification for parts of the zbMATH corpus~\cite{BarthelTB13}.
They criticized the micro averaged $F_1$ measure, especially, if the average is applied only to the best performing classes. However, they report a micro-averaged $F_1$ score of $67.1\%$ for the zbMATH corpus.
They suggested training classifiers for a precision of $95\%$ and assigning MSC class labels in a  semi-automated recommendation setup.
Moreover, they suggested to measure the human baseline (inter-annotator agreement) for the classification tasks.
Moreover, they found that the combination of mathematical expressions and textual features improves the $F_1$ score for certain MSC classes substantially.
In \citeyear{SchonebergS14}, \citeauthor{SchonebergS14}~\cite{SchonebergS14} implement a method that combined formulae and text using an adapted Part of Speech Tagging approach.
Their paper reported a sufficient precision of $>.75$, however, it did not state the recall.
The proposed method was implemented and is currently being used especially to pre-classify general journals~\cite{zbMATH06353861} with additional information, like references.
For a majority of journals, coarse- and fine-grained codes can be found by statistically analyzing the MSC codes from referenced documents matched within the zbMATH corpus.
The editor of zbMATH hypothesizes that the reference method outperforms the algorithm developed by \citeauthor{SchonebergS14}.
To confirm or reject this hypothesis was one motivation for this project.

The positive effect of mathematical features is confirmed by \citeauthor{SuzukiF17}~\cite{SuzukiF17}, who measured the classification performance based on an arXiv and mathoverflow dataset.
In contrast, \citeauthor{Scharpf2020}~\cite{Scharpf2020} could not measure a significant improvement of classification accuracy for the arxiv dataset when incorporating mathematical identifiers.
In their experiments \citeauthor{Scharpf2020} evaluated numerous machine learning methods, which extended~\cite{Evans17,SojkaNALS19} in terms of accuracy and run-time performance, and found that complex compute-intensive neural networks do not significantly improve the classification performance.

In this paper, we focus on the coarse-grained classification of the \emph{primary MSC subject number} (pMSCn) and explore how current machine learning approaches can be employed to automate this process.
In particular, we compare the current state of the art technology~\cite{Scharpf2020} with a part of speech (POS) preprocessing based system customized for the application in zbMATH from 2014~\cite{SchonebergS14}.
\\

\noindent\hspace{-.01\textwidth}\fbox{\parbox{1.02\textwidth}{%
We define the following research questions:
\begin{enumerate}
  \def\labelenumi{\arabic{enumi}.}
  \item
  Which evaluation metrics are most useful to assess the classifications? %
  \item
  Do mathematical formulae as part of the text improve the classifications?
  \item
  Does POS preprocessing~\cite{SchonebergS14} improve the accuracy of classifications?
  \item
  Which features are most important for accurate classification?
  \item
  How well do automated methods perform in comparison to a human baseline?
\end{enumerate}}}

\section{Method}\label{sec:method}
\begin{figure}[t]
    \centering
    \includegraphics[width=\textwidth]{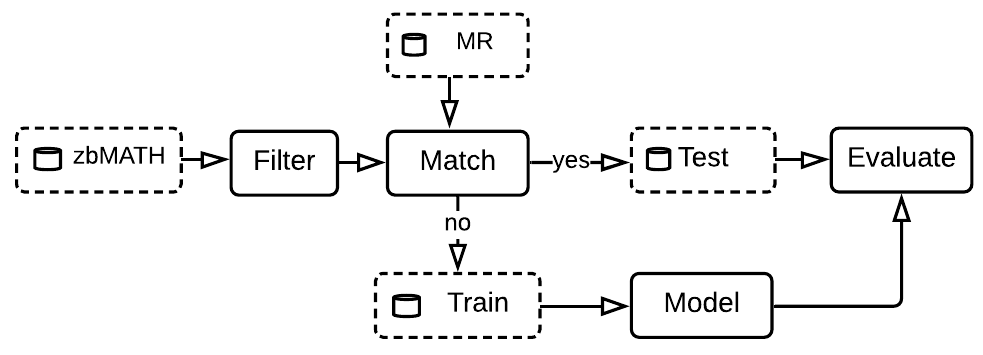}
    \caption{Workflow overview.}\label{fgWorkflow}
\end{figure}

To investigate the given set of problems, we first created test and training datasets.
We then investigated the different pMSCn encodings, trained our models and evaluated the results, cf Figure~\ref{fgWorkflow}.

\subsection{Generation of a test and training dataset}
\paragraph{Filter current high quality articles:}
The zbMATH database has assigned MSC codes to more than \href{https://zbmath.org/?q=cc\%3A*}{3.6 M} articles.
However, the way in which mathematical articles are written has changed over the last century, and the classification of historic articles is not something we aim to investigate in this article.
The first MSC was created in 1990, and has since been updated every ten years (2000, 2010, and 2020)~\cite{MSC2010}.
With each update, automated rewrite rules are applied to map the codes from the old MSC to the next MSC version, which is connected with a loss of accuracy of the class labels.
To obtain a coherent and high quality dataset for training and testing, we focused on the more recent articles from 2000 to 2019, which were classified using the MCS version 2010, and we only considered selected journals\footnote{The list of selected journals is available from \url{https://zbmath.org/?q=dt\%3Aj+st\%3Aj+py\%3A2000-2019.}}.
Additionally, we restricted our selection to English articles and limited ourselves to abstracts rather than reviews of articles.
To be able to compare methods that are based on references and methods using text and title, we only selected articles with at least one reference that could be matched to another article.
In addition, we excluded articles that were not yet published and processed.
The list of articles is available from our website: \url{https://automsceval.formulasearchengine.com}
\paragraph{Splitting to test and training set:}
After applying the filter criteria as mentioned above, we split the resulting list of 442,382 articles into test and training sets.
For the test set, we aimed to measure the bias of our zbMATH classification labels.
Therefore, we used the articles for which we knew the classification labels by the MR service as the training set from a previous research project~\cite{Bannister2018}.
The resulting test set consisted of \(n = 32,230\) articles, and the training set contained 410,152 articles.
To ensure that this selection did not introduce additional bias, we also computed the standard ten-fold cross validation, cf. Section 3.
\paragraph{Definition of article data format:}
To allow for reproducibility, we created a dedicated dataset from our article selection, which we aim to share with other researchers.
However, currently, legal restrictions apply and the dataset can not yet be provided for anonymous download at this date.
However, we can grant access for research purposes as done in the past~\cite{BarthelTB13}.
Each of the 442,382 articles in the dataset contained the following fields:
\begin{description}
  \item[de] An eight-digit ID of the document\footnotemark.
  \item[labels] The actual MSC codes\footnotemark[\value{footnote}].
  \item[title] The English title of the document, with LaTeX macros for mathematical language~\cite{Schubotz2019b}.
  \item[text] The text of the abstract with LaTeX macros.
  \item[mscs] A comma separated list of MSC codes generated from the references.
\end{description}
\footnotetext{The fields \texttt{de} and \texttt{labels} must not be used as input to the classification algorithm.}
These 5 fields were provided as CSV files to the algorithms. The \texttt{mscs} field was generated as follows: For each reference in the document, we looked up the MSC codes of the reference. For example, if a certain document contained the references \(A,B,C\) that are also in the documents in zbMATH and the MSC codes of \(A,B,C\) are \(a_{1}\) and \(a_{2}\), \(b_{1}\), and \(c_{1} - c_{3}\), respectively, then the field \texttt{mscs} will read \(a_{1}a_{2},b_{1},c_{1}c_{2}c_{3}.\)

After training, we required each of our tested algorithms to return the following fields in CSV format for the test sets:
\begin{description}
  \item[de (integer)] Eight-digit ID of the document.
  \item[method (char(5))] Five-letter ID of the run.
  \item[pos (integer)] Position in the result list.
  \item[coarse (integer)] Coarse-grained MSC subject number.
  \item[fine (char(5), optional)] Fine-grained MSC code.
  \item[score (numeric, optional)] Self-confidence of the algorithm about the result.
\end{description}
We ensured that the fields \texttt{de}, \texttt{method} and \texttt{pos} form a primary key, i.e., no two entries in the result can have the same combination of values.
Note that for the current multi-class classification problem, \texttt{pos} is always 1, since only the primary MSC subject number is considered.
\subsection{Definition of evaluation metrics}
While the assignment of all MSC codes to each article is a multi-label classification task, the assignment of the primary MSC subject, which we investigate in this paper, is only a multi-class classification problem.
With \(k = 63\) classes, the probability of randomly choosing the correct class of size \(c_{i}\) is rather low \(P_{i} = \frac{c_{i}}{n}.\)
Moreover, the dataset is not balanced.
In particular, the entropy \(
H = - \sum_{i = 1}^{k}P_{i}\log P_{i},
\) can be used to measure the imbalance \(\widehat{H} = \frac{H}{\log k}\) by normalizing it to the maximum entropy \(\log{k.}\)

To take into account the imbalance of the dataset, we used weighted versions of precision \(p\), recall \(r,\) and the \(F_{1}\) measure \(f\).
In particular, the precision \(p = \frac{\sum_{i = 1}^{k}c_{i}p_{i}}{n}\) with the class precision \(p_{i}\). \(r\) and \(F_{1}\) are defined analogously.

In the test set, no entries for the pMSCn 97 (Mathematics education) were included, thus

\[\widehat{H} = \frac{H}{\log k} = \frac{3.44}{\log 62} = .83\]

Moreover, we eliminate the effect of classes with only few samples by disregarding all classes with less than 200 entries.
While pMSCn with few samples have little effect on the average metrics, the individual values are distracting in plots and data tables.
Choosing 200 as the minimum evaluation class size reduces the number of effective classes to \(k = 37,\) which only has a minor effect on the normalized entropy as it is raised to \(\widehat{H} = .85.\)
The chosen value of 200 can be interactively adjusted in the dynamic result figures we made available online\footnote{\url{https://autoMSCeval.formulasearchengine.com}}.
Additionally, the individual values for \(P_{i}\) that were used to calculate \(H\) are given in the column \texttt{p} in the table on that page.
As one can experience in the online version of the figures, the impact on the choice of the minimum class size is insignificant.

\subsection{Selection of methods to evaluate}

In this paper, we compare 12 different methods for (automatically) determining the primary MSC subject in the test dataset:

\begin{description}
    \item[zb1] Reference MSC subject numbers from zbMATH.
  \item[mr1] Reference MSC subject numbers from MR.
  \item[titer]
  According to recent research performed on the arXiv dataset~\cite{Scharpf2020}, we chose a machine learning method with a good trade-off between speed and performance.
  We combined the \texttt{title}, abstract \texttt{text}, and reference \texttt{mscs} of the articles via string concatenation.
  We encoded these string sources using the \emph{TfidfVectorizer} of the Scikit-learn\footnote{\url{https://swmath.org/software/8058}~\cite{swSciKit}} python package.
  We did not alter the \emph{utf-8} encoding, and did not perform accent striping, or other character normalization methods, with the exception of lower-casing.
  Furthermore, we used the \emph{word} analyzer without a custom stop word list, selecting tokens of two or more alphanumeric characters, processing unigrams, and ignoring punctuation.
  The resulting vectors consisted of float64 entries with \emph{l2} norm unit output rows.
  This data was passed to Our encoder.
  The encoder was trained on the training set to subsequently transform or vectorize the sources from the test set.
  We chose a lightweight \emph{LogisticRegression} classifier from the python package Scikit-learn.
  We employed the \emph{l2} penalty norm with a \(10^{- 4}\) tolerance stopping criterion and a 1.0 regularization.
  Furthermore, we allowed intercept constant addition and scaling, but no class weight or custom random state seed.
  We fitted the classifier using the \emph{lbfgs} (Limited-memory BFGS) solver for 100 convergence iterations.
  These choices were made based on a previous study in which we clustered arXiv articles.
  \item[refs] Same as \texttt{titer}, but using only the \texttt{mscs} as input\footnotemark.
  \item[titls] Same as \texttt{titer}, but using only the \texttt{title} as input\footnotemark[\value{footnote}].
  \item[texts] Same as \texttt{titer}, but using only the \texttt{text} as input\footnotemark[\value{footnote}].
  \item[tite] Same as \texttt{titer}, but without using the \texttt{mscs} as input\footnotemark[\value{footnote}].
  \item[tiref]: Same as \texttt{titer}, but without using the abstract \texttt{text} as input \footnotemark[\value{footnote}].
  \item[teref]: Same as \texttt{titer}, but without using the \texttt{title} as input\footnotemark[\value{footnote}].
  \item[ref1] We used a simple SQL script to suggest the most frequent primary MSC subject based on the \texttt{mscs} input.
  This method is currently used in production to estimate the primary MSC subject.
  \item[uT1] 
  We adjusted the JAVA program posLingue~\cite{SchonebergS14} to read from the new training and test sets. However, we did not perform a new training and instead reused the model that was trained in 2014. However, for this run, we removed all mathematical formulae from the \texttt{title} and the abstract \texttt{text} to generate a baseline.
  \item[uM1]
  The same as uT1 but in this instance, we included the formulae.
  We slightly adjusted the formula detection mechanism, since the way in which formulae are written in zbMATH had changed~\cite{Schubotz2019b}.
  This method is currently used in production for articles that do not have references with resolvable \texttt{mscs}.
\end{description}

\footnotetext{Each of these sources was encoded and classified separately.}
\begin{table*}[t]
  \hfil
  \begin{tabular}{llll}
    \toprule
    {} &      p &      r &      f \\
    \midrule
    zb1   &      1 &      1 &      1 \\
    mr1   &  0.814 &  0.814 &  0.812 \\
    titer &  0.772 &  0.778 &  0.773 \\
    refs  &  0.748 &  0.753 &  0.746 \\
    titls &  0.637 &  0.627 &  0.623 \\
    texts &  0.699 &  0.709 &  0.699 \\
    ref1  &  0.693 &  0.648 &  0.652 \\
    uT1   &  0.656 &  0.642 &  0.645 \\
    uM1   &  0.655 &  0.639 &  0.644 \\
    tiref &   0.76 &  0.764 &   0.76 \\
    teref &  0.769 &  0.774 &   0.77 \\
    tite  &  0.713 &  0.722 &  0.713 \\
    \bottomrule
  \end{tabular}
  \hfil
  \begin{tabular}{llll}
    \toprule
    {} &      p &      r &      f \\
    \midrule
    zb1   &  0.817 &  0.807 &   0.81 \\
    mr1   &      1 &      1 &      1 \\
    titer &  0.776 &  0.775 &  0.772 \\
    refs  &  0.743 &  0.743 &  0.737 \\
    titls &  0.644 &  0.632 &  0.627 \\
    texts &  0.704 &  0.709 &  0.699 \\
    ref1  &  0.693 &  0.646 &  0.652 \\
    uT1   &  0.653 &  0.636 &  0.639 \\
    uM1   &  0.652 &  0.632 &  0.636 \\
    tiref &  0.762 &  0.761 &  0.758 \\
    teref &  0.771 &   0.77 &  0.767 \\
    tite  &   0.72 &  0.724 &  0.715 \\
    \bottomrule
  \end{tabular}
  \hfil
  \caption{Precision $p$, recall $r$ and $F_1$-measure $f$ with regard to
    the baseline \texttt{zb1} (left) and \texttt{mr1} (right).} \label{tb1}
\end{table*}
\section{Evaluation and Discussion}\label{sec:eval}

After executing each of the methods described in the previous section, we calculated the precision \(p\), recall \(r,\) and \(F_{1}\) score \(f\) for each method, cf. Table~\ref{tb1}.
Overall, we find that results are similar whether we used zbMATH or MR as a baseline in our evaluation.
Therefore, we will use zbMATH as the reference for the remainder of the paper. All data, including the test results using MR as the baseline is available from: \url{https://automsceval.formulasearchengine.com}.

\paragraph{Effect of mathematical expressions and part-of-speech tags:}
\begin{figure}[t]
  \centering
  \includegraphics[width=\textwidth]{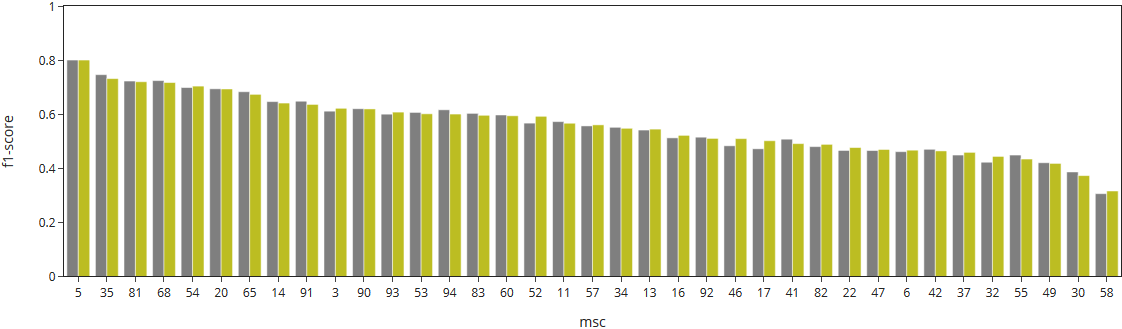}
  \caption{Mathematical symbols in \texttt{title} and abstract \texttt{text} do not improve the classification quality. Method \texttt{uT1} =left bar; method \texttt{uM1}=right bar%
 }\label{fgMath}
  \includegraphics[width=\textwidth]{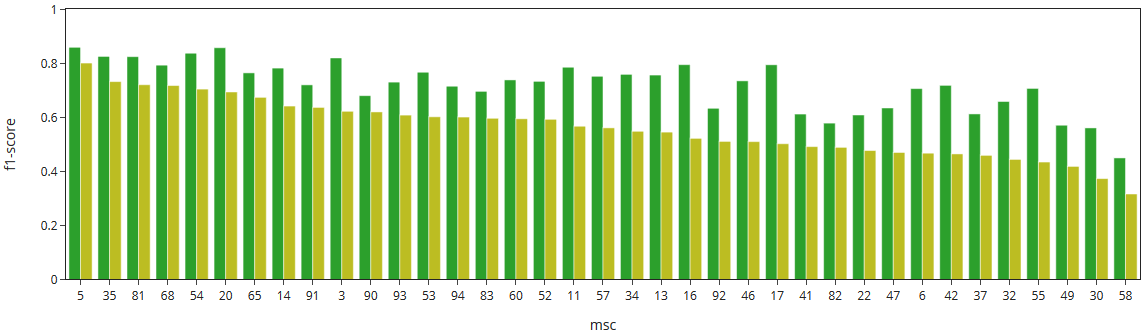}
\caption{Part-of-speech tagging for mathematics does not improve the classification quality. Method \texttt{tite}=left bar, method \texttt{uM1}=right bar.}\label{fgPOS}
\end{figure}
By filtering out all mathematical expressions in the current production method \texttt{uT1} in contrast to \texttt{uM1} we could receive information on the impact of mathematical expressions on classification quality.
We found that the overall \(F_{1}\) score without mathematical expressions \(f_{uT1} = 64.5\%\) is slightly higher than the score with mathematical expressions \(f_{uM1} = 64.4\%.\)
Here, the main effect is an increase in recall from \(63.9\%\) to \(64.2\%.\)
Additionally, a class-wise investigation showed that for most classes, \texttt{uT1} outperformed \texttt{uM1}, cf. Figure~\ref{fgMath}.
Exceptions are pMSCn 46 (Functional analysis ) and 17 (Nonassociative rings and algebras) where the inclusion of math tags raised the \(F_{1}\)-score slightly.

We evaluated the effect of \emph{part of speech tagging} (POS), by comparing \texttt{tite} with \texttt{uM1}. \(f_{\mathrm{tite}} = .713\) clearly outperformed \(f_{uM1} = .64.\)
This held true for all MSC subjects, cf. Figure~\ref{fgPOS}.
We modified posLingo to output the POS tagged text and used this text as input and retrained scikit learn classifier \texttt{tite2}.
However, this method did not lead to better results than \texttt{tite}.

\begin{figure}[t]
  \centering
  \includegraphics[width=1\textwidth]{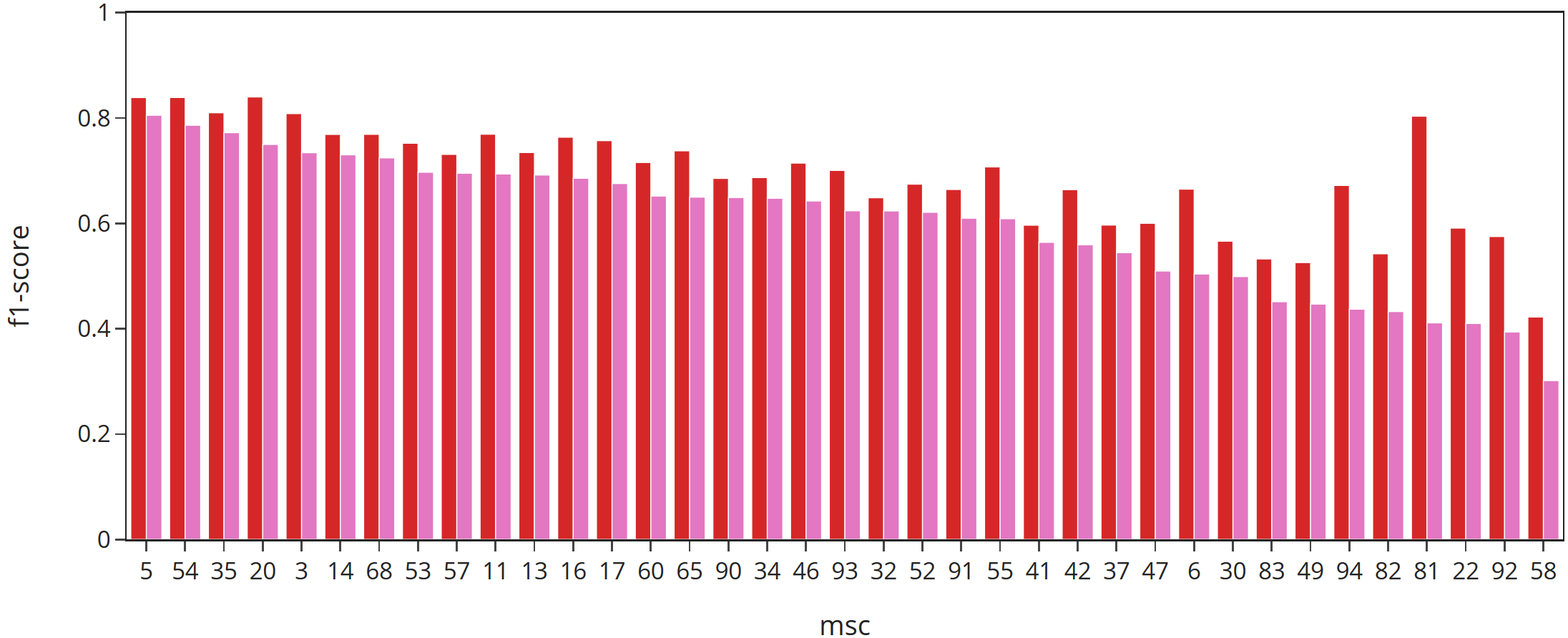}
  \caption{Machine learning method (\texttt{refs}, left)  clearly outperforms current production (\texttt{ref1}, right) method using references as only source for classification.}\label{fgRefs}
\end{figure}

\paragraph{Effect of features and human baseline:}
The newly developed method combined method~\cite{Scharpf2020} works best in a combined approach that uses \texttt{title}, abstract \texttt{text}, and references \texttt{titer} \(f_{\mathrm{titer}} = 77.3\%.\)
This method performs significantly better than methods that omit either one of these features.
The best performing single feature method was \texttt{refs} \(f_{\mathrm{refs}} = 74.6\%\)) followed by \texttt{text} \(f_{\mathrm{text}} = 69.9\%\) and \texttt{titls} \(f_{\mathrm{titls}} = 62.3\%\).
Thus, automatically generating the MSC subject while including the references appears to be a very valuable strategy.
This becomes evident also when comparing the scores of approaches that only considered two features. For the approaches that excluded \texttt{title} (i.e. \texttt{teref} \(f_{\mathrm{text}} = 77\%\)) or abstract \texttt{text}\texttt{\ }(i.e.\texttt{\ }\texttt{tiref} \(f_{\mathrm{text}} = 76\%\)), the performance remained notably higher than when the approach excluded the reference \texttt{mscs}\ (\texttt{tite} \(f_{\mathrm{text}} = 71.3\%\))
However, it is also worth pointing out that the naive reference-based method, \texttt{ref1} \(f_{\mathrm{text}} = 65.2\%\), which is currently being used in production still performs more poorly than just using \texttt{tite} despite this approach ignoring references.
In conclusion, we can say that training a machine learning algorithm that weights all information from the fine grained MSC codes is clearly better than the majority vote of the references, cf. \ref{fgRefs}.

\begin{figure}[t]
  \includegraphics[width=1\textwidth]{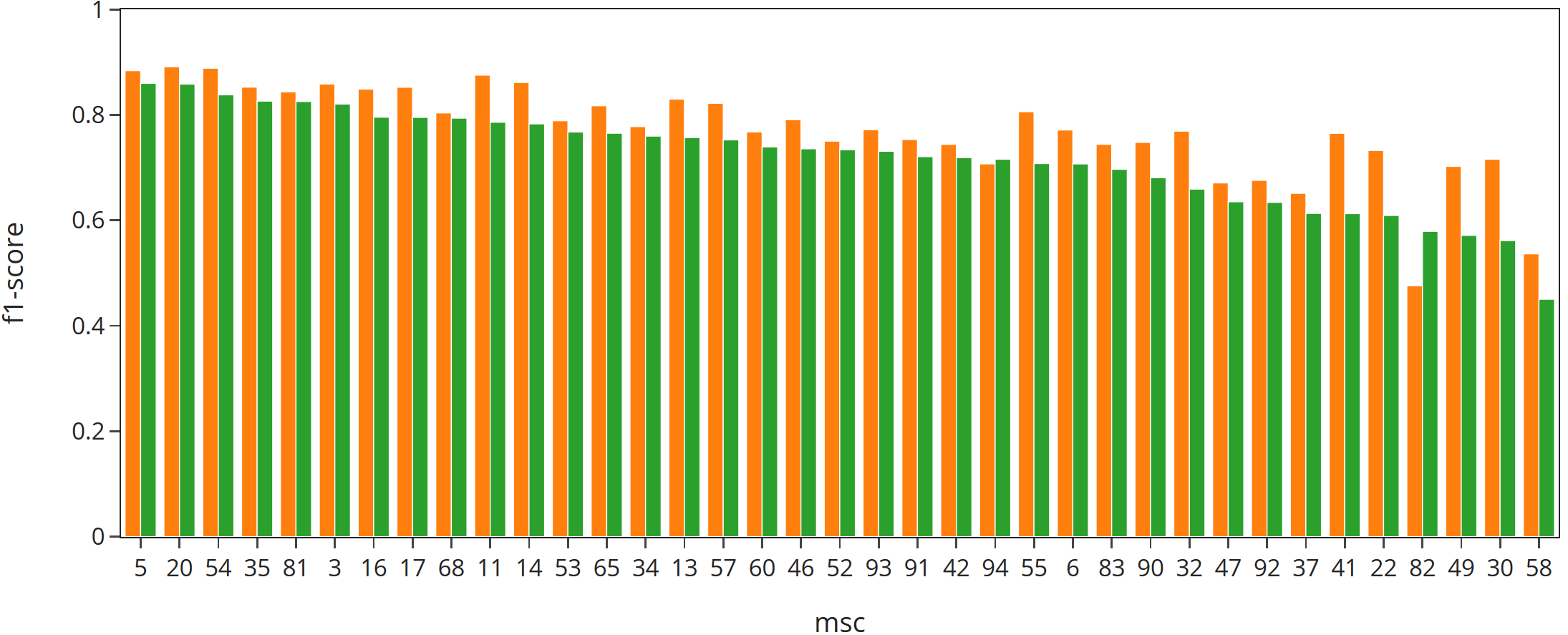}
  \caption{For many pMSCn the best automatic method (\texttt{titer}, right) gets close to the performance of the human baseline (\texttt{mr1} left)}\label{fgHum}
\end{figure}
Even the best performing machine learning algorithm, \texttt{titer} with \(f_{\mathrm{titer}} = 77.3\%\), is worth than using the classification by human experts from MR, the other mathematics publication reviewing service, resulted in a baseline of \texttt{mr1} \(f_{mr1} = 81.2\%.\)
However, there is no foundation that could allow us to determine which of the primary MSC subjects, either from MR or zbMATH, are truly correct.
Assigning a two-digit label to mathematical research papers -- which often cover overlapping themes and topics within mathematics -- remains a challenge even to humans, who struggle to conclusively label publications as belonging to only a single class.
While for some classes, expert agreement is very high, e.g. for class 20 agreement is \(89.1\%\), for other classes, such as 82, agreement is only at \(47.6\%\) regarding the \(F_{1}\) score, cf., Figure~\ref{fgHum}.
These discrepancies reflect the intrinsic problem that mathematics cannot be fully reflected by a hierarchical system.
The differences in classifications made among the two reviewing services are likely also a reflection of emphasizing different facets of evolving research, which often derive from differences in the reviewing culture.

We also investigated the bias introduced by the non-random selection of the training set.
Performing ten fold cross validation on the entire dataset yielded an accuracy of \(f_{titer,10} = .776\) with a standard deviation \(\sigma_{titer,10} = .002.\)
Thus, test set selection does not introduce a significant bias.

\begin{figure}[t]
  \centering
  \includegraphics[width=\textwidth]{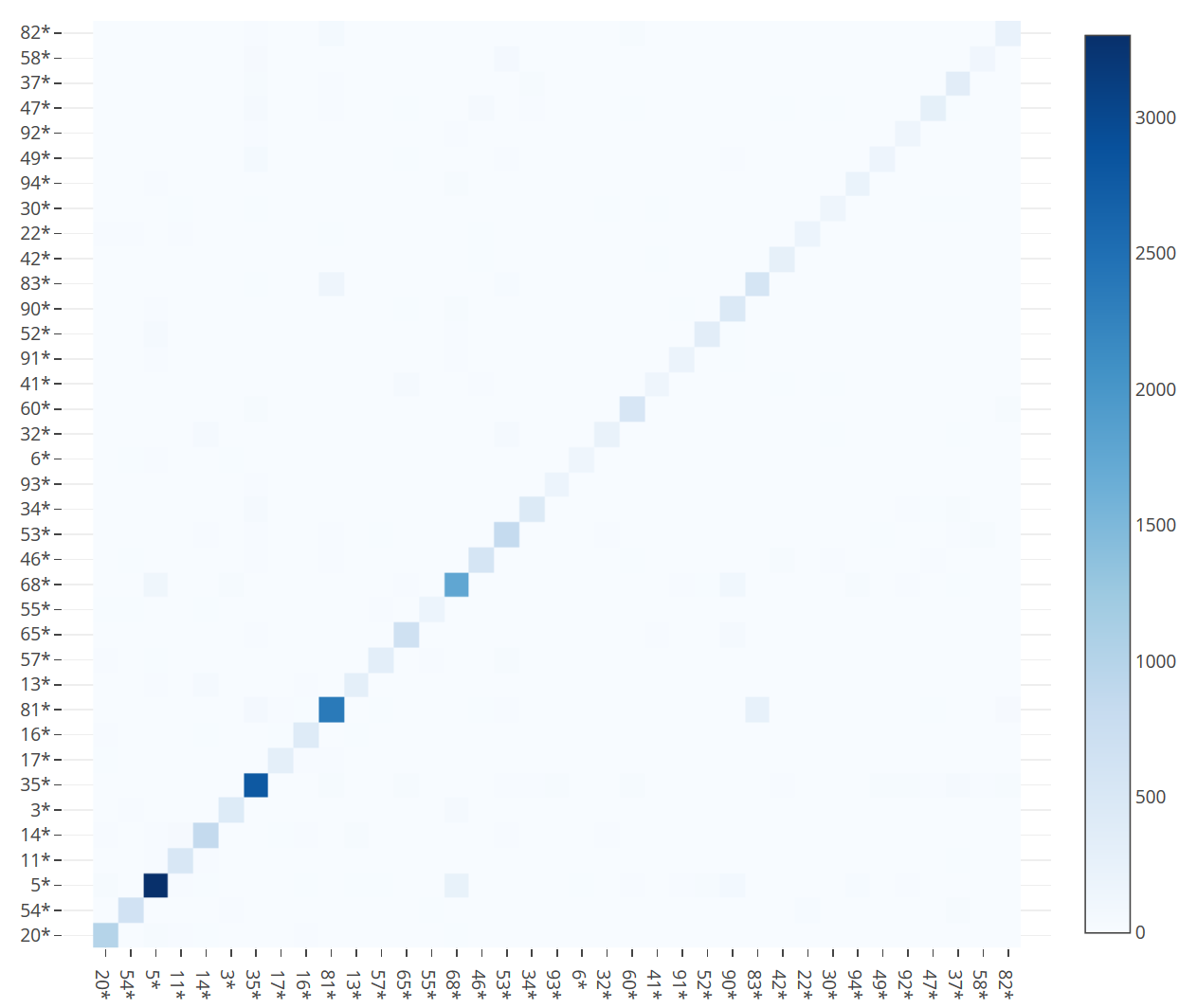}
  \caption{Confusion matrix \texttt{titer}}\label{fgConfusion}
\end{figure}
\begin{figure}[t]
  \centering
  \includegraphics[width=\textwidth]{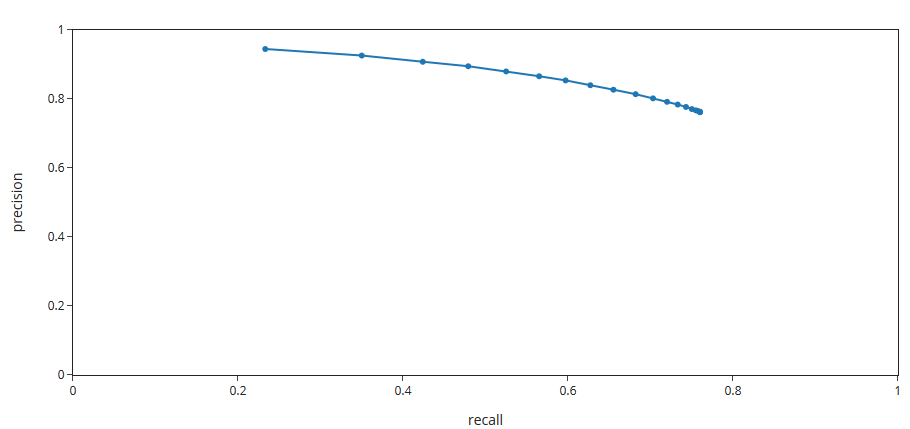}
  \caption{Precision recall curve \texttt{titer}.}
  \label{fgPR}
\end{figure}

After having discussed the strengths and weaknesses of the individual methods tested, we now discuss how the currently best-performing method, \texttt{titer}, can be improved.
One standard tool to analyze misclassifications is a confusion matrix, cf., Figure~\ref{fgConfusion}.
In this matrix, off-diagonal elements of the matrix indicate that two sets of classes are often mixed by the classification algorithm.
The x axis shows the true labels, while the y axis shows the predicted labels.
The most frequent error of \emph{titer} was that 68 (Computer science) was classified as 5 (Combinatorics).
Moreover, 81 (Quantum theory) and 83 (Relativity and gravitational theory) were often mixed up.

However, in general the number of misclassifications were small and there was no immediate action that one could take to avoid special cases of misclassification that do not involve a human expert.

\begin{figure}[t]
  \centering
  \includegraphics[width=0.56\textwidth]{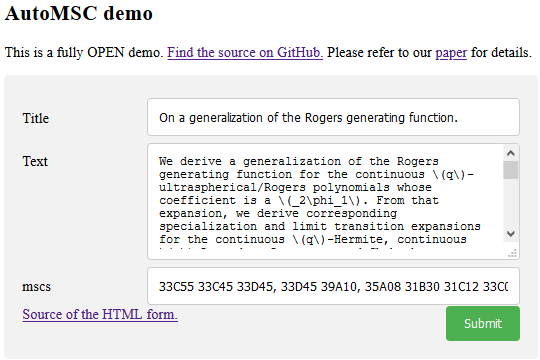}
  \includegraphics[width=0.42\textwidth]{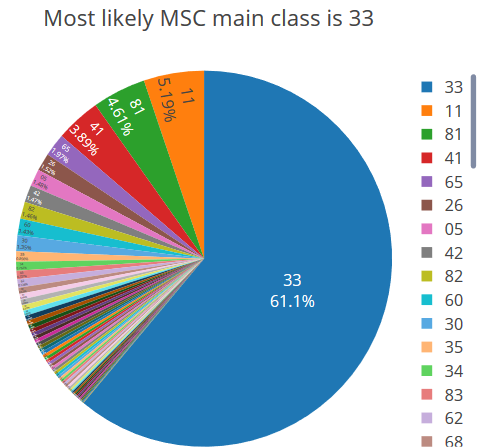}
  \caption{Classification frontend}\label{fgScreenshot}
\end{figure}

Since \texttt{titer} outperforms both the text-based and reference based methods currently used in zbMATH, we decided to develop a restful API that wraps our trained model into a service.
We use pythons fastAPI under unicorn to handle higher loads.
Our system is available as a docker container and can thus be scaled on demand.
To simplify development and testing, we provide a static HTML page as a micro UI, which we call \emph{AutoMSC}.
This UI displays not only lists/suggests the most likely primary MSC subjects but also the less likely MSC subjects.
We expect that our UI can support human experts, especially whenever the most likely MSC subject seems unsuitable.
The result is displayed as a pie-chart, cf., Figure~\ref{fgScreenshot} from \url{https://automscbackend.formulasearchengine.com}.
To use the system in practice, an interface to the citation matching component of zbMATH would be desired to paste the actual references rather than the MSC subjects extracted from the references.
Moreover, looking at the precision-recall curve (Figure~\ref{fgPR}) for \texttt{titer}, suggests that one can also select a threshold for falling back to manual classification.
For instance, if one requires a precision that is as high as the precision of the other human classifications by MR, one would need to only consider suggestions with a score \(> 0.5\).
This would automatically classify \(86.2\%\) of the 135k articles being annually classified by subject experts at zbMATH/MR and thus significantly reduce the number of articles that humans must manually examine without a loss of classification quality.
This is something we might develop in the future.

\section{Conclusion \& Future Work}\label{sec.concl}
Returning to our research questions, we summarize our findings as follows:
First, we asked which metrics are best suited to assess classification quality.
We demonstrated that the classification quality for the primary MSC subject can be evaluated with classical information retrieval methods such as precision, recall and \(F_{1}\)-score.
We share the observation \citeauthor{BarthelTB13}~\cite{BarthelTB13} that the averages do not reflect the performance of outliers, cf. Figures~1-4.
However, for our methods the difference between the best and worst performing class was significantly smaller than reported by~\cite{BarthelTB13}.

Second, we wanted to find out whether taking into account the mathematical formulae contained in publications could improve the accuracy of classifications.
In accordance with~\cite{Scharpf2020}, we did not find evidence that mathematical expressions improved pMSCn classification.
However, we did not evaluate advanced encodings of mathematical formulae. This is will be a subject of future work, cf. Figure~1.

Third we evaluated the effect of POS-preprocessing~\cite{SchonebergS14} and found that modern machine learning methods do not benefit from the POS tagging based model developed by \cite{SchonebergS14}, cf. Figure~2.

Fourth we evaluated which features are most important for an accurate classification.
We conclude that references have the highest prediction power, followed by the abstract text and title.

Finally, we evaluated the performance of automatic methods in comparison to a human baseline.
We found that our best performing method has an $F_1$ score of 77.2\%.
The manual classification is significantly better for most classes, cf. Figure 4.
However, the self-reported \texttt{score} can be used to reduce the manual classification effort by 86.2\%, without a loss in classification quality.

In the future, we plan to extend our automated methods to predict full MSC codes.
Moreover, we would like to be able to assign pMSCn to document sections, since we realize that some research just does not fit into one of the classes.
We also plan to extend the application domain to other mathematical research artifacts, such as blog posts, software, or dataset descriptions.
As a next step, we plan to generate pMSCn from authors using the same methods we applied for references.
We speculate that authors will have a high impact on the classification, since authors often publish in the same field.
For this purpose, we are leveraging our prior research on affiliation disambiguation, which could be used as fallback method for junior authors, who have not yet established a track record.
Another extension is a better combination of the different features.
Especially when performing research on the full MSC code-generation, we will need to use a different encoding for the MSC from references and authors.
However, this new encoding requires more main memory for the training of the model and cannot be done on a standard laptop.
Thereafter, we will re-investigate the impact of mathematical formulae since the inherently combined representation of text and formulae was not successful.

Our work represents a further step in the automation of Mathematics Subject Classification and can thus support reviewing services, such as \emph{zbMATH} or \emph{Mathematical Reviews}.
For accessible exploration, we have made the best-performing approaches available in our \emph{AutoMSC} implementation and have shared our code on our website.
We envision that other application domains requiring an accurate labeling of publications into their respective Mathematics Subject Classification, for example, research paper recommendation systems, or reviewer recommendation systems, will also be able to benefit from this work.
AutoMSC delivers comparable results to human experts in the first stage of MSC labeling, all without requiring manual labor or trained experts.
In the future, zbMATH will use our new method for all journals that used to employ the method by~\citeauthor{SchonebergS14}~\cite{SchonebergS14} introduced in 2014.

\paragraph{Acknowledgments:} This work was supported by the German Research Foundation (DFG grant GI 1259-1).
The authors would like to express their gratitude to Felix Hamborg, and Terry Ruas for their advice in the most recent machine learning technology.
\printbibliography[keyword=primary]